\documentclass[12pt]{article}
\usepackage{amssymb}
\makeatletter
\def\numberbysection{\@addtoreset{equation}{section}
 	\def\theequation{\thesection.\arabic{equation}}}
\makeatother

\numberbysection

\newcommand{\be}{\begin{eqnarray}}
\newcommand{\ee}{\end{eqnarray}}
\newcommand{\non}{\nonumber}
\newcommand{\rr}{\rangle}
\newcommand{\ket}[1]{|{#1}\rangle}
\newcommand{\dket}[1]{|{#1}\rangle\hskip -3pt\rangle}
\newcommand{\Z}{\ensuremath{\mathsf{Z}}}

\newcommand{\R}[3]{\renewcommand{\arraystretch}{0.5}
R\!\begin{array}{l}{\ \ \scriptstyle {#1}}
\\ {\scriptstyle{#2}}\\ 
{\ \ \scriptstyle{#3}}\end{array}
\renewcommand{\arraystretch}{1.0}}

\begin{document}

\begin{titlepage}
\strut\hfill UMTG--234
\vspace{.5in}
\begin{center}

\LARGE Supersymmetry in the boundary tricritical Ising field theory\\[1.0in]
\large Rafael I. Nepomechie\\[0.8in]
\large Physics Department, P.O. Box 248046, University of Miami\\[0.2in]  
\large Coral Gables, FL 33124 USA\\

\end{center}

\vspace{.5in}

\begin{abstract}
We argue that it is possible to maintain both supersymmetry and
integrability in the boundary tricritical Ising field theory.  Indeed,
we find two sets of boundary conditions and corresponding boundary
perturbations which are both supersymmetric and integrable.  The first
set corresponds to a ``direct sum'' of two non-supersymmetric theories
studied earlier by Chim.  The second set corresponds to a one-parameter
deformation of another theory studied by Chim.  For both cases, the
conserved supersymmetry charges are linear combinations of $Q$, $\bar
Q$ and the spin-reversal operator $\Gamma$.
\end{abstract}

\end{titlepage}

\setcounter{footnote}{0}

\section{Introduction}\label{sec:intro}

Consider a $1+1$-dimensional quantum field theory which is integrable
in the bulk.  Demanding that this field theory remain integrable in
the presence of a boundary places severe restrictions on the possible
boundary interactions \cite{GZ}.  Similarly, if a field theory has
bulk supersymmetry, then demanding that supersymmetry be preserved in
the presence of a boundary evidently also restricts the possible
boundary interactions.  These considerations immediately raise the
question: to what extent can {\it both} integrability and
supersymmetry be maintained in the presence of a boundary?  We address
this question here in the context of the tricritical Ising field
theory \cite{Za1} -- i.e., the tricritical Ising conformal field 
theory (CFT) \cite{BPZ, CFT, SCFT} perturbed by the $\Phi_{(1\,, 3)}$ 
operator \cite{Za2}. Several authors have already investigated this 
field theory in the presence of a boundary \cite{Ch}-\cite{FPR}. 
Although the question of whether supersymmetry can be maintained in
the boundary theory was raised in the seminal work of Chim \cite{Ch},
it has not been addressed until now.  \footnote{An alternative field
theory for perturbed tricritical Ising has been proposed \cite{Fe,
DMM}, which we shall not discuss here.}

We have recently investigated the issue of integrability and
supersymmetry in the presence of a boundary for other models: the
scaling supersymmetric Yang-Lee model \cite{AN}, and the $N= 1$ and
$N=2$ sine-Gordon models \cite{Ne1}.  However, in contrast to those
models whose particles have vertex-type scattering matrices, the
tricritical Ising field theory contains kinks which have RSOS-type
scattering matrices.  Also, the tricritical Ising field theory is an
example of a perturbed minimal model which, unlike the supersymmetric
Yang-Lee model, is unitary.

To the above question, we find an affirmative answer: it {\it is}
possible to maintain both supersymmetry and integrability in the
boundary tricritical Ising field theory.  Indeed, we find two sets of
boundary conditions and corresponding boundary perturbations which are
both supersymmetric and integrable.  The first boundary condition
involves a superposition of two pure ``Cardy'' boundary conditions
\cite{Ca}.  Hence, the corresponding field theory is in fact a
``direct sum'' of two non-supersymmetric theories studied in
\cite{Ch}.  We explicitly construct the conserved supersymmetry
charge, and find that it contains a term proportional to the
spin-reversal operator.  The field theory corresponding to the second
set of boundary conditions is a one-parameter deformation of another
theory studied in \cite{Ch}.

In Section \ref{sec:bulk}, we briefly review the pertinent results
from \cite{Za1} on the bulk tricritical Ising field theory.  We also
introduce the spin-reversal operator $\Gamma$, which -- as already
noted -- plays an important role in the boundary theory.  In Section
\ref{sec:CBC}, we recall \cite{Ch} the conformal boundary conditions
and corresponding conformal boundary states of the tricritical Ising
CFT, and we argue that certain boundary states and combinations thereof
have superconformal symmetry. Our main results are contained in 
Sections \ref{sec:NS} and \ref{sec:R}, where we study supersymmetric
perturbations of these superconformal boundary conditions.  In
particular, we propose specific boundary perturbing operators and the
corresponding conserved supersymmetry charges and boundary $S$
matrices.  We conclude in Section \ref{sec:disc} with a brief
discussion of our results.

\section{Bulk TIM}\label{sec:bulk}

Zamolodchikov has described \cite{Za1} the bulk massive integrable
quantum field theory obtained as a perturbation of the tricritical
Ising CFT by the $\Phi_{(1\,, 3)}$ operator.  This model, to which we
shall refer as the (bulk) tricritical Ising field theory, or simply
(bulk) TIM, can be described by the ``action''
\be
A = A_{{\cal M}(4/5)} + \lambda \int_{-\infty}^{\infty} dy 
\int_{-\infty}^{\infty} dx\  
\Phi_{({3\over 5}\,, {3\over 5})}(x \,, y) \,, \qquad \lambda < 0 \,, 
\label{bulkaction}
\ee
where $A_{{\cal M}(4/5)}$ is the action of the tricritical Ising CFT
(i.e., the minimal unitary model ${\cal M}(4/5)$ with central charge 
$c={7\over 10}$), and $\Phi_{({3\over 5}\,,{3\over 5})}$ is the spinless 
$(1\,, 3)$ primary field of this CFT with dimensions 
$({3\over 5} \,, {3\over 5})$. The ${\cal M}(4/5)$ Kac table is 
given in Table \ref{figM45}.
\begin{table}[htb] 
  \centering
  \begin{tabular}{|c|c|c|c|}\hline
    $\frac{3}{2}$ & $\frac{3}{5}$ & $\frac{1}{10}$ & 0 \\
    \hline
    $\frac{7}{16}$ & $\frac{3}{80}$ & $\frac{3}{80}$ & $\frac{7}{16}$ \\
    \hline
    0 & $\frac{1}{10}$ & $\frac{3}{5}$ & $\frac{3}{2}$ \\
    \hline
   \end{tabular}
  \caption{Kac table for ${\cal M}(4/5)$}
  \label{figM45}
\end{table}
Moreover, $\lambda$ is a bulk parameter with dimension 
length${}^{-{4\over 5}}$.
Following \cite{Za1}, we restrict our attention to the case $\lambda < 
0$, for which there is a three-fold vacuum degeneracy, and the 
spectrum consists of massive kinks $K_{a \,, b}(\theta)$ that separate 
neighboring vacua, $a \,, b \in \{ -1 \,, 0 \,, 1\}$ with $|a - b|=1$. 
Multi-kink states 
\be
| K_{a_{1} \,, b_{1}}(\theta_{1})\ 
K_{a_{2} \,, b_{2}}(\theta_{2}) \ldots \rr 
\ee 
must obey the adjacency conditions $b_{1} = a_{2}\,, $ etc. 

The two-kink $S$ matrix has four distinct amplitudes defined by
\cite{Za1}
\be
| K_{0\,, a}(\theta_{1})\ K_{a\,, 0}(\theta_{2}) \rr_{in} &=&
A_{0}(\theta_{12})| K_{0\,, a}(\theta_{2})\ K_{a\,, 0}(\theta_{1}) 
\rr_{out} +
A_{1}(\theta_{12})| K_{0\,, -a}(\theta_{2})\ K_{-a\,, 0}(\theta_{1}) 
\rr_{out} \,, \non \\
| K_{a\,, 0}(\theta_{1})\ K_{0\,, a}(\theta_{2}) \rr_{in} &=&
B_{0}(\theta_{12})| K_{a\,, 0}(\theta_{2})\ K_{0\,, a}(\theta_{1}) 
\rr_{out} \,,\non \\
| K_{a\,, 0}(\theta_{1})\ K_{0\,, -a}(\theta_{2}) \rr_{in} &=&
B_{1}(\theta_{12})| K_{a\,, 0}(\theta_{2})\ K_{0\,, -a}(\theta_{1}) 
\rr_{out} \,,
\label{bulkS}
\ee
where $\theta_{12}= \theta_{1} - \theta_{2}$, and $a=\pm 1$.
These amplitudes are given by \cite{Za1}
\be
A_{0}(\theta) &=&  \cosh {\theta\over 4}\ A(\theta) 
\,, \qquad 
A_{1}(\theta) = -i \sinh {\theta\over 4}\ A(\theta)
\,, \non \\
B_{0}(\theta) &=&  \cosh {1\over 4}(\theta - i \pi)\ B(\theta) 
\,, \qquad 
B_{1}(\theta) =  \cosh {1\over 4}(\theta + i \pi)\ B(\theta) \,, 
\label{bulkamplitudes}
\ee
where 
\be
A(\theta) = e^{\gamma \theta} S(\theta) \,, \qquad 
B(\theta) = \sqrt{2} e^{-\gamma \theta} S(\theta) \,,
\ee
with $\gamma = {1\over 2 \pi i}\ln 2$, and 
\be
S(\theta) = {1\over \sqrt{\pi}}\prod_{k=1}^{\infty} 
{\Gamma(k - {\theta\over 2 \pi i}) 
\Gamma(-{1\over 2}+ k + {\theta\over 2 \pi i})
\over
\Gamma({1\over 2}+ k - {\theta\over 2 \pi i}) 
\Gamma(k + {\theta\over 2 \pi i})}
\,.
\label{bulkscalarfactor}
\ee 
As remarked in \cite{Za1}, these amplitudes do not have poles in the
physical strip, corresponding to the fact that there are no bulk bound
states.  Also, these amplitudes are essentially the Boltzmann weights
of the critical Ising lattice model \cite{Ba, ABF}.

The tricritical Ising CFT contains a conserved supercurrent, and in
fact has superconformal symmetry \cite{SCFT}.  The holomorphic
component $G(z)$ of the supercurrent corresponds to the primary field
$\Phi_{({3\over 2}\,, 0)}$ with spin ${3\over 2}$; and similarly, the
antiholomorphic component $\bar G(z)$ is the primary field
$\Phi_{(0\,, {3\over 2})}$ with spin $-{3\over 2}$.  
Using the methods developed in \cite{Za2}, 
Zamolodchikov has shown that in the perturbed theory
(\ref{bulkaction}) the supercurrent remains conserved,
\be
\partial_{\bar z} G = \partial_{z} \bar \Psi \,, \qquad
\partial_{z} \bar G = \partial_{\bar z} \Psi \,,
\ee
where $\bar \Psi = 4 \lambda \bar G_{-{1\over 2}} \Phi_{({1\over 10}\,, 
{1\over 10})}$ and 
$\Psi = 4 \lambda G_{-{1\over 2}} \Phi_{({1\over 10}\,, {1\over 10})}$. 
Hence, the model has fermionic integrals
of motion $Q$ and $\bar Q$ of spin $\pm {1\over 2}$, respectively,
\be
Q = \int G\ dz + \bar \Psi\ d\bar z \,, \qquad
\bar Q = \int \bar G\ d\bar z +  \Psi\ dz \,,
\label{bulkSUSYcharges}
\ee 
that obey the $N=1$ supersymmetry algebra
\be
Q^{2} = P \,, \qquad \bar Q^{2} = \bar P \,, \qquad \{ Q \,, \bar Q 
\} = 2 t \,,
\ee
where $t$ is the topological charge.  The action of the supersymmetry
charges on multi-kink states (both ``in'' and ``out'') is given 
by \cite{Za1}
\be
\lefteqn{Q | K_{a_{1}\,, a_{2}}(\theta_{1})\ 
K_{a_{2}\,, a_{3}}(\theta_{2}) 
\ldots K_{a_{N}\,, a_{N+1}}(\theta_{N}) \rr 
= \sum_{j=1}^{N} \Bigl\{ \sqrt{m}\ \beta(a_{j} \,, a_{j+1})\ 
e^{{\theta_{j}\over 2}}} \non  \\ 
&\times & | K_{-a_{1}\,, -a_{2}}(\theta_{1}) \ldots 
K_{-a_{j-1}\,, -a_{j}}(\theta_{j-1})\ K_{-a_{j}\,, 
a_{j+1}}(\theta_{j})\ \ldots
K_{a_{N}\,, a_{N+1}}(\theta_{N}) \rr \Bigr\} \,,
\label{Qaction}
\ee
and 
\be
\lefteqn{\bar Q | K_{a_{1}\,, a_{2}}(\theta_{1})\ 
K_{a_{2}\,, a_{3}}(\theta_{2}) 
\ldots K_{a_{N}\,, a_{N+1}}(\theta_{N}) \rr 
= \sum_{j=1}^{N} \Bigl\{ \sqrt{m}\ \bar \beta(a_{j} \,, a_{j+1})\ 
e^{-{\theta_{j}\over 2}}} \non  \\ 
&\times & | K_{-a_{1}\,, -a_{2}}(\theta_{1}) \ldots 
K_{-a_{j-1}\,, -a_{j}}(\theta_{j-1})\ K_{-a_{j}\,, 
a_{j+1}}(\theta_{j})\ \ldots
K_{a_{N}\,, a_{N+1}}(\theta_{N}) \rr \Bigr\} \,,
\label{barQaction}
\ee
where
\be
\beta( a\,, b) = i (a+ i b) \,, \qquad 
\bar \beta( a\,, b) = -i (a- i b) \,,
\label{betas}
\ee
and $m$ is the kink mass. Moreover, the topological 
charge acts according to 
\be
t | K_{a_{1}\,, a_{2}}(\theta_{1}) \ldots
K_{a_{N}\,, a_{N+1}}(\theta_{N}) \rr = -(a_{1}^{2} - a_{N+1}^{2})
| K_{a_{1}\,, a_{2}}(\theta_{1}) \ldots
K_{a_{N}\,, a_{N+1}}(\theta_{N}) \rr \,.
\ee
One can show that these charges commute 
with the above $S$ matrix. Indeed, this is how the $S$ matrix is 
determined in \cite{Za1}.

We now define the spin-reversal operator $\Gamma$ by the following 
action on  multi-kink states (both ``in'' and ``out''):
\be
\lefteqn{\Gamma\ | K_{a_{1}\,, a_{2}}(\theta_{1})
\ K_{a_{2}\,, a_{3}}(\theta_{2}) 
\ldots K_{a_{N}\,, a_{N+1}}(\theta_{N}) \rr} \non \\
& & =
| K_{-a_{1}\,, -a_{2}}(\theta_{1})\ K_{-a_{2}\,, -a_{3}}(\theta_{2}) 
\ldots K_{-a_{N}\,, -a_{N+1}}(\theta_{N}) \rr \,.
\label{spinreversal}
\ee
Evidently, the spin-reversal operator satisfies
\be
\Gamma^{2} = 1 \,,
\ee
and it commutes with the bulk $S$ matrix (\ref{bulkS}). Moreover, $\Gamma$ 
anticommutes with the supersymmetry charges,
\be
\{ \Gamma \,, Q \} = 0 \,, \qquad \{ \Gamma \,, \bar Q \} = 0 \,,
\label{anticommutes}
\ee
as follows from the fact $\beta(-a \,, -b) = -\beta(a \,, b)$, and 
similarly for $\bar \beta(a \,, b)$.
These properties suggest that $\Gamma$ corresponds in the (perturbed)
CFT to the operator $(-1)^{F}$, where $F$ is the Fermion-number
operator.

\section{Conformal boundary conditions}\label{sec:CBC}

The Cardy states \cite{Ca} for the tricritical Ising boundary CFT are
given in terms of Ishibashi states \cite{Is} by \cite{Ch}
\be
(-) &:& \quad
\ket {\widetilde{0}} = C \left[ \left(\dket 0 + \dket {3\over 2} \right)
+ \eta \left(  \dket {1\over 10}  + \dket {3\over 5} \right) 
+ \sqrt[4]{2}\dket {7\over 16} 
+  \sqrt[4]{2}\dket {3\over80} \right] \,, \non  \\
(-0) &:&\quad
\ket {\widetilde{1\over 10}} = C \left[ \eta^2 \left( \dket 0 
+ \dket {3\over 2} \right) 
- \eta^{-1}\left( \dket{1\over 10} + \dket {3\over 5}\right) 
-\sqrt[4]{2} \eta^2 \dket {7\over{16}}
+ \sqrt[4]{2}\eta^{-1} \dket {3\over{80}} \right] \,, \non \\
(0+) &:& \quad
\ket {\widetilde{3\over 5}} = C \left[ \eta^2 \left( \dket  0 
+ \dket {3\over 2}\right)
- \eta^{-1} \left( \dket {1\over{10}} +\dket {3\over 5} \right)
+ \sqrt[4]{2}\eta^2 \dket {7\over 16}
- \sqrt[4]{2}\eta^{-1} \dket {3\over 80} \right]\,, \non \\
(+) &:& \quad
\ket {\widetilde {3\over 2}} = C \left[ \left( \dket 0 
+ \dket {3\over 2} \right)
+ \eta \left( \dket {1\over 10} + \dket {3\over 5} \right)
- \sqrt[4]{2} \dket {7\over 16} 
- \sqrt[4]{2} \dket {3\over 80} \right]\,, \non \\
(0) &:& \quad
\ket {\widetilde{7\over 16}} = \sqrt{2} C \left[ \left(\dket 0 
-\dket {3\over2}\right)
- \eta \left( \dket {1\over 10} - \dket {3\over 5}\right) \right] \,, \non \\
(d) &:& \quad
\ket {\widetilde{3\over 80}} = \sqrt{2} C \left[ \eta^2 \left( \dket 0 
- \dket {3\over 2} \right)
+ \eta^{-1} \left( \dket {1\over 10} - \dket {3\over 5}\right) \right]  \,,
\label{Cardystates}
\ee
where 
\be
C = \sqrt{{\sin{\pi\over 5}}\over{\sqrt{5}}} \,, \qquad 
\eta = \sqrt{{\sin{{2\pi}\over 5}}\over{\sin{\pi\over 5}}} \,.
\ee
In (\ref{Cardystates}) are also given the corresponding conformal
boundary conditions (CBC) which Chim has identified.  Let us recall
that, in the bulk, the three vacua $-1 \,, 0 \,, +1$ are degenerate. 
However, these vacua do not necessarily remain degenerate at the
boundary.  Indeed, for the boundary conditions $(-) \,, (0) \,, (+)$,
the order parameter is fixed at the boundary to the vacua $-1 \,, 0
\,, +1$, respectively.  For the boundary condition $(-0)$, the vacua
$-1$ and $0$ are degenerate at the boundary; hence, the order
parameter at the boundary may be in either of these two vacua. 
Similarly, for the boundary condition $(0+)$, the $0$ and $+1$ vacua
are degenerate at the boundary.  Finally, for the boundary condition
$(d)$, all three vacua $-1 \,, 0 \,, +1$ are degenerate at the
boundary (as well as in the bulk); i.e., the order parameter at the
boundary may be in any of the three vacua.

In the remainder of this Section, we argue that the boundary states
corresponding to the conformal boundary conditions $(-) \& (+)$, $(-0)
\& (0+)$, $(o)$ and $(d)$ have superconformal symmetry.  We observe
that the boundary states corresponding to these conformal boundary
conditions are given by
\be
(-) \& (+) &:&  \quad 
\ket {\widetilde{0}} + \ket {\widetilde {3\over 2}} =
2C \left( \dket {0_{+}} + \eta \dket {{1\over 10}_{+}} \right)
\,, \non  \\
(-0) \& (0+) &:& \quad 
\ket {\widetilde{1\over 10}} + \ket {\widetilde{3\over 5}} = 
2C \left( \eta^2 \dket {0_{+}} - \eta^{-1} \dket {{1\over 10}_{+}} \right)
\,, \non  \\
(o) &:& \quad 
\ket {\widetilde{7\over 16}} = \sqrt{2} C \left(
\dket {0_{-}} - \eta \dket {{1\over 10}_{-}} \right)
\,, \non \\
(d) &:& \quad 
\ket {\widetilde{3\over 80}} = \sqrt{2} C \left(
\eta^2 \dket {0_{-}} + \eta^{-1} \dket {{1\over 10}_{-}} \right)  \,,
\label{superstates}
\ee
where
\be
\dket {0_{\pm}} =  \dket 0 \pm \dket {3\over 2}  \,, \qquad 
\dket {{1\over 10}_{\pm}} = \dket {1\over 10} \pm \dket {3\over 5} \,.
\label{plusminus}
\ee
Notice that the first two states in (\ref{superstates}) are 
superpositions of ``pure'' Cardy states. \footnote{For 
recent discussions of superpositions of Cardy states, see e.g. 
\cite{GRW, RRS, Gr}.} Also, these states are related by duality 
\cite{Ch}. Indeed, under the duality transformation ${\cal D}$, 
the Ishibashi states $\dket 0$ and $\dket {3\over 5}$ remain invariant,
while the states $\dket {1\over 10}$ and $\dket {3\over 2}$ pick up a 
minus sign. (See Appendix \ref{sec:appendix}.) Thus,
\be
{\cal D} \dket {0_{\pm}} = \dket {0_{\mp}} \,, \qquad 
{\cal D} \dket {{1\over 10}_{\pm}} = - \dket {{1\over 10}_{\mp}} \,.
\ee
It follows that
\be
{\cal D} \left( 
\ket {\widetilde{0}} + \ket {\widetilde {3\over 2}}
\right) = \sqrt{2} \ket {\widetilde{7\over 16}} \,, \qquad 
{\cal D} \left(  
\ket {\widetilde{1\over 10}} + \ket {\widetilde{3\over 5}}
\right) = \sqrt{2} \ket {\widetilde{3\over 80}} \,.
\label{duality}
\ee 

Let us briefly review some basic facts about superconformal field
theory \cite{SCFT}.  The $N=1$ superconformal algebra is defined by
the (anti) commutation relations
\be
\left[ L_{m} \,, L_{n} \right] &=& (m-n) L_{m+n} + {1\over 12}c  
(m^{3}-m) \delta_{m+n \,, 0} \,, \non \\
\left[ L_{m} \,, G_{r} \right] &=&  ({1\over 2}m - r) G_{m+r} 
\,, \non \\
\left\{ G_{r} \,, G_{s} \right\} &=& 2 L_{r+s} + {1\over 3} c (r^{2} - 
{1\over 4}) \delta_{r+s \,, 0} \,,
\label{algebra}
\ee
where $r \,, s \in \Z$ for the Ramond ($R$) sector and $r \,, s \in \Z +
{1\over 2}$ for the Neveu-Schwarz ($NS$) sector. The operators 
$\bar L_{n}$, $\bar G_{r}$ corresponding to the antiholomorphic 
components of the energy-momentum tensor and supercurrent obey similar
relations.  Highest weight irreducible representations of the
holomorphic algebra are generated from highest weight states $\ket
\Delta$ satisfying
\be
L_{0} \ket \Delta  = \Delta \ket \Delta  \,, \qquad 
L_{n} \ket \Delta = G_{r} \ket \Delta  = 0 \,, 
\quad n > 0 \,, \quad r > 0 \,.
\label{highestweights}
\ee
The corresponding highest weight states of the antiholomorphic algebra 
are denoted by $\ket {\bar \Delta}$.
The tricritical Ising model, which has $c={7\over 10}$, can be regarded 
either as the conformal minimal model ${\cal M}(4/5)$ or the 
superconformal minimal model ${\cal SM}(3/5)$.

Since the Ishibashi states $\dket j$ are annihilated by $(L_{n} - \bar
L_{-n})$, then so are the linear combinations (\ref{superstates}).  We
wish to show that \footnote{If $G_{r}$ and $\bar G_{r}$ are assumed to
be operators in the ``open'' channel, then the rotation to the
``closed'' channel produces certain $\pm i$ factors. For convenience,
we shall assume here instead that the operators $G_{r}$ and $\bar G_{r}$
obeying (\ref{algebra}) are in the closed channel.}
\be
\left( G_{r} - \bar G_{-r} \right) 
\left( \ket {\widetilde{1\over 10}} + \ket {\widetilde{3\over 5}} 
\right) &=& 0 \,, \qquad 
\left( G_{r} - \bar G_{-r} \right) 
\left( \ket {\widetilde{0}} + \ket {\widetilde{3\over 2}} 
\right) = 0 \,, \non \\
\left( G_{r} + \bar G_{-r} \right) \ket {\widetilde{7\over 16}}  &=& 0 \,,
\qquad 
\left( G_{r} + \bar G_{-r} \right) \ket {\widetilde{3\over 80}}  = 0 \,,
\label{want}
\ee
for $r \in Z + {1\over 2}$. 
In view of (\ref{superstates}), it suffices to show that the boundary
states $\dket {0_{\pm}}$, $\dket {{1\over 10}_{\pm}}$ obey
\be
\left( G_{r} \mp \bar G_{-r} \right) \dket {0_{\pm}} = 0 \,, \qquad
\left( G_{r} \mp \bar G_{-r} \right) \dket {{1\over 10}_{\pm}}  = 0 \,.
\label{sufficient}
\ee
(This result is stated without proof in \cite{Ne2}.)  To this end, we
observe that the vacuum state $\ket 0$ of ${\cal M}(4/5)$ and ${\cal
SM}(3/5)$ is the same, since it is unique.  Thus, it satisfies
\be
L_{n} \ket 0 &=& 0 \,, \qquad n \ge -1 \,, \non \\ 
G_{r} \ket 0 &=& 0 \,, \qquad r \ge -{1\over 2} \,. 
\label{vacuum}
\ee
The state $\ket {1\over 10}$, which is a highest weight in both 
${\cal M}(4/5)$ and ${\cal SM}(3/5)$, satisfies
\be
G_{r}\ket {1\over 10} = 0 \,, \qquad r > 0 \,.
\label{onetenth}
\ee 
The state $\ket {3\over 2}$, which is a highest weight in ${\cal 
M}(4/5)$, is a descendant in ${\cal SM}(3/5)$,
\be
\ket {3\over 2} = \sqrt{15\over 7} G_{-{3\over 2}} \ket 0 \,,
\label{threehalves}
\ee
where the coefficient is fixed by the normalization condition
$\langle {3\over 2} \ket  {3\over 2} = 1$. Similarly,
\be
\ket {3\over 5} = \sqrt{5} G_{-{1\over 2}} \ket {1\over 10} \,.
\label{threefifth}
\ee
With the help of the representation for the conformal Ishibashi states 
given in \cite{DRTW}
\be
\dket j &=& \left(1 + {L_{-1} \bar L_{-1}\over 2 \Delta_{j}} + \ldots 
\right) \ket j \otimes \ket {\bar j} \,, \qquad  j \ne 0 \,, \non \\
\dket 0 &=& \left(1 + {L_{-2} \bar L_{-2}\over c/2} + \ldots 
\right) \ket 0 \otimes \ket {\bar 0} \,,
\ee
one can proceed to verify (\ref{sufficient}). Indeed,
for the case $r={1\over 2}$,
\be
\lefteqn{\left( G_{1\over 2} - \bar G_{-{1\over 2}} \right)  \dket {0_{+}} =
\left( G_{1\over 2} - \bar G_{-{1\over 2}} \right) \Bigl( 
\ket 0 \otimes \ket {\bar 0} + \ket {3\over 2} \otimes \ket {\bar {3\over 2}}} 
\non \\
& & + {20\over 7} L_{-2} \ket 0 \otimes \bar L_{-2} \ket {\bar 0} 
+ {1\over 3} L_{-1} \ket {3\over 2} \otimes \bar L_{-1} \ket {\bar {3\over 2}} 
+ \ldots \Bigr) \,.
\label{example1}
\ee 
The operator $\left( G_{1\over 2} - \bar G_{-{1\over 2}} \right)$ 
annihilates the first term due to (\ref{vacuum}); and acting on the 
second term, this operator gives
\be
-2 \sqrt{15\over 7} \ket {3\over 2} \otimes \bar L_{-2}\ket {\bar 0} \,,
\ee
as follows from (\ref{threehalves}), (\ref{algebra}) and
(\ref{vacuum}).  There is an equal and opposite contribution from the
action of $G_{1\over 2}$ on the third term in (\ref{example1}).  The
actions of $\bar G_{-{1\over 2}}$ on the third term, and of $\left(
G_{1\over 2} - \bar G_{-{1\over 2}} \right)$ on the fourth term,
produce contributions which presumably are cancelled by corresponding
contributions from higher-order terms represented by ellipses in
(\ref{example1}). Similarly,
\be
\left( G_{1\over 2} - \bar G_{-{1\over 2}} \right)  
\dket {{1\over 10}_{+}} &=&
\left( G_{1\over 2} - \bar G_{-{1\over 2}} \right) \Bigl( 
\ket {1\over 10} \otimes\ket  {\bar {1\over 10}} 
+ \ket {3\over 5} \otimes \ket {\bar {3\over 5}} + \ldots \Bigr) \non \\
&=& G_{1\over 2} \ket {1\over 10} \otimes \ket {\bar {1\over 10}} -
\ket {1\over 10} \otimes \bar G_{-{1\over 2}}\ket {\bar {1\over 10}} +
 G_{1\over 2} \ket {3\over 5} \otimes \ket {\bar {3\over 5}} + \ldots
\label{example2}
\,.
\ee 
The first term vanishes due to (\ref{onetenth}). The second term
is equal to
\be
-{1\over \sqrt{5}}\ket {1\over 10} \otimes \ket {\bar {3\over 5}} \,,
\ee
as follows from (\ref{threefifth}); and it is cancelled by the third
term.  Other values of $r$ can presumably be treated in a similar
manner.

\section{Boundary TIM: NS case}\label{sec:NS}

We shall consider supersymmetric perturbations of the tricritical
Ising boundary CFT with two different (super)conformal boundary
conditions.  In this Section we consider the CBC $(-0) \& (0+)$; and
in Section \ref{sec:R} we consider the CBC $(d)$.  We refer to these
two cases as NS and R, respectively, since these are the sectors to 
which the corresponding boundary states belong.

\subsection{Definition of the model as a perturbed CFT}\label{sec:PCFT}

We now consider the boundary tricritical Ising field theory, with
the action \footnote{This boundary action differs in two 
important respects from a similar one considered by Chim \cite{Ch}: 
(1) While he considers the CBC $(-0)$ corresponding 
to a pure Cardy state, we consider the CBC $(-0) \& (0+)$ 
corresponding to a superposition state. As we have already argued, the 
latter is supersymmetric, while the former is not. (2) Chim considers a 
single perturbing boundary operator ($\phi_{({3\over 5})\,, (-0)}$ in our 
notation), whereas we consider the difference of two such operators. 
The latter generates an RG boundary flow which is supersymmetric, while the 
former does not.} 
\be
A &=& A_{{\cal M}(4/5)+ (-0) \& (0+)} + \lambda \int_{-\infty}^{\infty} dy 
\int_{-\infty}^{0} dx\  
\Phi_{({3\over 5}\,, {3\over 5})}(x \,, y) \non \\
&-& h \int_{-\infty}^{\infty} dy\  
\left( \phi_{({3\over 5})\,, (-0)}(y) - \phi_{({3\over 5})\,, (0+)}(y) \right) 
\,, \label{boundaryaction}
\ee
where again we restrict to the case $\lambda < 0$. We now explain each 
of the terms in turn.

The first term in (\ref{boundaryaction}) is the action for the
tricritical Ising boundary CFT ${\cal M}(4/5)$ with the conformal
boundary condition $(-0) \& (0+)$.  
We have argued in the previous Section that the conformal boundary state 
corresponding to this CBC is annihilated not only by
$(L_{n} - \bar L_{-n})$, but also by $(G_{r} - \bar G_{-r})$. Hence, it 
is in fact a superconformal boundary state. The corresponding 
boundary condition is superconformal, 
\be
\left( T(y+ix) - \bar T(y-ix) \right) \Big\vert_{x=0} = 0 \,, \qquad 
\left( G(y+ix) - \bar G(y-ix) \right) \Big\vert_{x=0} = 0 \,.
\label{SCBC1}
\ee

The second term in the action (\ref{boundaryaction}) is the bulk
perturbation, which is the same as in the bulk action
(\ref{bulkaction}), except that the $x$ integral is now restricted to
the half-line ${x \le 0}$.

The last term in the action (\ref{boundaryaction}) is the boundary
perturbation. It involves the boundary primary fields 
$\phi_{({3\over 5})\,, (-0)}$ and $\phi_{({3\over 5})\,, (0+)}$ with dimension 
$\Delta_{(1\,, 3)} = {3\over 5}$ which act on $(-0)$ and $(0+)$, 
respectively. \footnote{In general, boundary operators $\phi_{a}$ and 
$\phi_{b}$ which act on conformal boundary conditions $a$ and $b$ commute; 
i.e., their operator product expansion with each other is zero.
Such operators have recently been studied in \cite{Gr}.} The
reason for the relative minus sign between the two boundary primary 
fields will be given below, when we discuss boundary flows.
Moreover, $h$ is a boundary parameter which has
dimensions length${}^{-{2\over 5}}$. 

Since the boundary perturbation has the same dimension ($\Delta_{(1\,,
3)}$) as the bulk perturbation, the analysis of \cite{GZ} suggests
that this boundary perturbation is integrable.  One can also use the
arguments of \cite{GZ} to infer that the boundary perturbation
preserves supersymmetry.  Indeed, consider the bulk conformal limit
$\lambda = 0$. In view of the boundary condition (\ref{SCBC1}),
by computing the operator product
$\left[ G(y+ix) - \bar G(y-ix) \right] 
\left( \phi_{({3\over 5})\,, (-0)}(y') - \phi_{({3\over 5})\,, (0+)}(y') 
\right)$,
one can conclude to first order in perturbation theory that the quantity
\be
\hat Q = \int_{-\infty}^{0} dx\ \left[ G(x \,, y) + \bar G(x \,, y) \right] 
+ \Theta(y) \,,
\ee
with $\Theta(y) \propto h (1 - 2 \Delta_{(1\,, 3)})
\left( \phi_{({1\over 10})\,, (-0)}(y) - \phi_{({1\over 10})\,, (0+)}(y) 
\right)$ 
is an integral of motion. It is plausible that, for the general 
massive case $\lambda \ne 0$, this becomes
\be
\hat Q = Q + \bar Q + \Theta \,,
\label{boundSUSY}
\ee
where $Q$ and $\bar Q$ are  given by (\ref{bulkSUSYcharges}) 
with the $x$ integrals restricted to the half-line.

Various arguments \cite{Ch, LSS, Af, GRW, FPR, RRS} support the following 
pattern of renormalization group boundary flows for the tricritical 
Ising boundary CFT:
\be
(-0) + \phi_{({3\over 5})} & \rightarrow &  (0) \,, \qquad 
(-0) - \phi_{({3\over 5})} \rightarrow (-) \,, \label{nonSUSYflows1} \\
(0+) + \phi_{({3\over 5})} & \rightarrow & (+) \,, \qquad 
(0+) - \phi_{({3\over 5})} \rightarrow (0) \,. \label{nonSUSYflows2}
\ee 
This would imply that the boundary perturbation in
(\ref{boundaryaction}) generates the boundary flows
\be
(-0) \& (0+) & \rightarrow & 2(0)  
\qquad \qquad \mbox{for} \qquad h<0 
\,, \non \\
(-0) \& (0+) & \rightarrow & (-) \& (+) 
\qquad \mbox{for} \qquad h>0 \,.
\label{SUSYflows1}
\ee
We have argued in Section \ref{sec:CBC} that both the pure Cardy state 
$\ket {\widetilde{7\over 16}}$ which corresponds to the CBC $(0)$, 
and also the superposition state
$\ket {\widetilde {0}} + \ket {\widetilde {3\over 2}}$ which 
corresponds to the CBC $(-) \& (+)$, are superconformal boundary 
states. Hence, the boundary flows (\ref{SUSYflows1}) connect
superconformal boundary conditions, and we refer to such flows as 
``supersymmetric flows.'' Indeed, we arrived at the particular 
boundary perturbation in (\ref{boundaryaction}) (specifically, 
the relative minus sign between the boundary primary fields 
$\phi_{({3\over 5})\,, (-0)}$ and $\phi_{({3\over 5})\,, (0+)}$) by 
requiring that it produce this supersymmetric flow.  Curiously, since
the $g$ factors \cite{AL} satisfy the relations
\be
g_{(-0)} = g_{(0+)} \,, \qquad g_{(-)} = g_{(+)} \,,
\ee
the ratio $g_{UV}/g_{IR}$ of $g$ factors corresponding to the
ultraviolet and infrared fixed points is the same for the
non-supersymmetric flows (\ref{nonSUSYflows1}),(\ref{nonSUSYflows2}) 
and for the corresponding supersymmetric flows (\ref{SUSYflows1}).
That is,
\be
{g_{(-0)}\over g_{(0)}} &=& {g_{(0+)}\over g_{(0)}}
= {g_{(-0)\& (0+)}\over g_{2(0)}}\,, 
\non \\ 
{g_{(-0)}\over g_{(-)}} &=& {g_{(0+)}\over g_{(+)}}
= {g_{(-0)\& (0+)}\over g_{(-)\& (+)}} \,.
\ee 

It is important to notice that the boundary perturbation in
(\ref{boundaryaction}) breaks spin-reversal symmetry.  Indeed, under
spin reversal, $(-0) \leftrightarrow (0+)$.  Hence, although the CBC
$(-0) \& (0+)$ remains invariant, the perturbation
$\left( \phi_{(-0)} - \phi_{(0+)} \right)$ picks up a minus sign.

\subsection{Boundary scattering theory}\label{sec:scattering}

We now turn to the boundary scattering theory.  Following Chim
\cite{Ch}, we assume that the boundary can have (at most) three
possible states, corresponding to the three different vacua.
We therefore define the boundary operator $B_{a}$ with $a 
\in \{ -1 \,, 0 \,, 1\}$, in terms of which the possible boundary 
states are $| B_{a} \rr$, with corresponding energies $e_{a}$.
Multi-kink states have the form
\be
| K_{a_{1} \,, a_{2}}(\theta_{1})\ 
K_{a_{2} \,, a_{3}}(\theta_{2}) \ldots 
K_{a_{N} \,, a}(\theta_{N})\ B_{a}\rr 
\ee
We extend the action (\ref{Qaction}), (\ref{barQaction}) of the
supercharges $Q$ and $\bar Q$ on such states in the obvious way
(namely, $B_{a}$ remains invariant);
and we extend the action (\ref{spinreversal}) of the spin-reversal
operator $\Gamma$ such that $\Gamma\ B_{a} = B_{-a}$.

The kink boundary $S$ matrix has six amplitudes defined by \cite{Ch}
\be
| K_{1\,, 0}(\theta)\ B_{0} \rr_{in} &=&
R_{+}(\theta)| K_{1\,, 0}(-\theta)\ B_{0} \rr_{out} \,, \non \\
| K_{-1\,, 0}(\theta)\ B_{0} \rr_{in} &=&
R_{-}(\theta)| K_{-1\,, 0}(-\theta)\ B_{0} \rr_{out} \,, \non \\
| K_{0\,, 1}(\theta)\ B_{1} \rr_{in} &=&
P_{+}(\theta)| K_{0\,, 1}(-\theta)\ B_{1} \rr_{out} 
+ V_{+}(\theta)| K_{0\,, -1}(-\theta)\ B_{-1} \rr_{out} 
\,, \non \\
| K_{0\,, -1}(\theta)\ B_{-1} \rr_{in} &=&
P_{-}(\theta)| K_{0\,, -1}(-\theta)\ B_{-1} \rr_{out} 
+ V_{-}(\theta)| K_{0\,, 1}(-\theta)\ B_{1} \rr_{out}  \,.
\label{boundS}
\ee

The unitarity constraints are given by \cite{Ch}
\be
R_{+}(\theta) R_{+}(-\theta) &=& 1 \,, \non  \\ 
P_{+}(\theta) V_{+}(-\theta) + V_{+}(\theta) P_{-}(-\theta)  &=& 0 
\,, \non  \\ 
P_{+}(\theta) P_{+}(-\theta) + V_{+}(\theta) V_{-}(-\theta)  &=& 1 
\label{unitarity}
\,,
\ee
together with the equations obtained by interchanging 
$+ \leftrightarrow -$; and the cross-unitarity constraints \cite{GZ}
are given by \cite{Ch}
\be
R_{+}({i \pi\over 2} - \theta) &=& 
A_{0}(2 \theta) R_{+}({i \pi\over 2} + \theta) + 
A_{1}(2 \theta) R_{-}({i \pi\over 2} + \theta) \,, \non  \\ 
P_{+}({i \pi\over 2} - \theta) &=& 
B_{0}(2 \theta) P_{+}({i \pi\over 2} + \theta) \,, \non  \\ 
V_{+}({i \pi\over 2} - \theta) &=& 
B_{1}(2 \theta) V_{+}({i \pi\over 2} + \theta) \,,
\label{crossunitarity}
\ee
together with the equations obtained by interchanging 
$+ \leftrightarrow -$.

Motivated by the plausibility argument given above that the boundary 
TIM (\ref{boundaryaction}) has a conserved supersymmetry charge
(\ref{boundSUSY}) and by the observation that the boundary
perturbation breaks spin-reversal symmetry (see also \cite{AN}, 
\cite{Ne1}), we consider the following operator
\be
\hat Q = Q + \bar Q + {2 i \sqrt{m}\over \alpha} \Gamma \,,
\label{hatQ1}
\ee
where $\alpha$ is a parameter which is yet to be determined. 
Requiring that this operator commute with the boundary $S$ matrix
(\ref{boundS}) yields the following constraints on the amplitudes:
\be
{R_{+}(\theta)\over R_{-}(\theta)} &=& 
{1 + \alpha \sinh {\theta\over 2}\over 1 - \alpha \sinh {\theta\over 2}}
\,, \non  \\
V_{+}(\theta) &=&  V_{-}(\theta) \equiv V(\theta) \,, \non  \\
V(\theta) &=&  {i\over 2 \alpha \cosh {\theta\over 2}}
\left( P_{-}(\theta)-P_{+}(\theta) \right)\,.
\label{SUSYconstraints1}
\ee
These constraints are a special case of those obtained in \cite{Ch} 
from the boundary Yang-Baxter equations.

Our goal is to determine the boundary $S$ matrix for the perturbation
of the CBC $(-0) \& (0+)$.  To this end, it is important to first
recall \cite{Ch} some results for the simpler (non-supersymmetric)
case of the perturbation of the CBC $(-0)$. For that case, the boundary 
can exist in either the states (vacua) $-1$ or $0$. The 
energies $e_{-1}$ and $e_{0}$ of these states depend on the 
value of the boundary parameter $h$: for $h=0$, the two energies
are equal; while for nonzero $h$, the energies are no longer degenerate. 
Consider the situation
(say, positive $h$) that $e_{-1} < e_{0}$; that is, the state
$-1$ is the ground state of the boundary, and the state $0$ is an 
excited state of the boundary. Since the boundary cannot exist in the 
state $+1$, the amplitudes $P_{+}(\theta)$ and $V_{\pm}(\theta)$ vanish. 
\footnote{Such amplitudes evidently violate the third unitarity 
constraint in (\ref{unitarity}). However, since the boundary
cannot exist in the state $+1$, this constraint should not be imposed.}
The amplitude $P_{-}(\theta)$ is given by $P_{-}(\theta) = P(\theta)$, 
where 
\be
P(\theta) = P_{\xi}^{CDD}(\theta)\  P_{min}(\theta) \,,
\label{Ptheta}
\ee
where $P_{min}(\theta)$ is the minimal solution of the equations
\be
P_{min}(\theta) P_{min}(-\theta) = 1 \,, \qquad 
P_{min}({i \pi\over 2} - \theta) = 
B_{0}(2 \theta) P_{min}({i \pi\over 2} + \theta) \,,
\ee
with no poles in the physical strip, and is given by \cite{Ch} 
\be
P_{min}(\theta) = e^{\gamma \theta} \prod_{k=1}^{\infty}
{\Gamma( k - {\theta\over 2 \pi i})^{2} 
\Gamma(k - {1\over 4} + {\theta\over 2 \pi i})
\Gamma(k + {1\over 4} + {\theta\over 2 \pi i})\over 
\Gamma( k + {\theta\over 2 \pi i})^{2} 
\Gamma(k + {1\over 4} - {\theta\over 2 \pi i})
\Gamma(k - {1\over 4} - {\theta\over 2 \pi i})} \,. 
\label{Pmin}
\ee 
Moreover, $P_{\xi}^{CDD}(\theta)$ is the CDD factor
\be
P_{\xi}^{CDD}(\theta) = 
{\sin \xi - i \sinh \theta \over \sin \xi + i \sinh \theta} \,,
\label{Pcdd}
\ee
which has a pole at $\theta = i \xi$.  The parameter $\xi$ is related 
in some way to the boundary parameter $h$. The state $0$, which is an 
excited state of the boundary, can be regarded as a boundary bound state,
which is associated with this pole when it lies in the physical strip
$0 \le \xi \le {\pi\over 2}$. The energies of the states $-1$ and $0$
are related by
\be
e_{0} = e_{-1} + m \cos \xi \,.
\ee
In particular, $h=0$ corresponds to $\xi = {\pi\over 2}$.
Using the boundary bound state bootstrap equations \cite{GZ}, one can
determine \cite{Ch} the amplitudes $R_{\pm}(\theta)$ ,
\be
R_{\pm}(\theta) = {1\over 2}(\cos {\xi\over 2} \pm i \sinh {\theta\over 2})
B(\theta - i \xi) B(\theta + i \xi) P(\theta) \,.
\label{Rpm}
\ee
For the opposite sign of $h$, the situation is reversed: $e_{0} <
e_{-1}$, and so the state $0$ is the ground state of the boundary, and
the state $-1$ is an excited state of the boundary.  Chim has
explained in detail how the above boundary $S$ matrix can give rise to
the RG boundary flows in Eq.  (\ref{nonSUSYflows1}).

For the case of the perturbation of the CBC $(0+)$, the results are 
parallel: the boundary can exist in either the states 
$0$ or $+1$. Hence, the amplitudes $P_{-}(\theta)$ and $V_{\pm}(\theta)$ 
vanish,  $P_{+}(\theta) = P(\theta)$, and $R_{\pm}(\theta)$ is given 
by (\ref{Rpm}). The corresponding RG boundary flows are given in 
Eq. (\ref{nonSUSYflows2}).

Finally, let us return to the case of the perturbation of the CBC
$(-0) \& (0+)$.  Since the corresponding boundary state is a
superposition of two ``pure'' boundary states, the vacua $-1$ and $+1$
are not states of the same ``irreducible'' theory.  Indeed, we have a
``direct sum'' of two ``irreducible'' theories: one theory with only boundary
states $-1$ and $0$, and another theory with only boundary states $0$
and $+1$.  In particular, the unitarity, crossing and bootstrap
constraints involving both $-1$ and $+1$ boundary states should not be
imposed.  Thus, the boundary $S$ matrix is the ``direct sum'' of the
boundary $S$ matrices given above for the perturbations of $(-0)$ and
$(0+)$.  That is, we propose for the boundary TIM
(\ref{boundaryaction}) the following boundary $S$ matrix:
\be
P_{+}(\theta) &=& P_{-}(\theta) = P(\theta)  \,, \qquad 
V_{\pm}(\theta) = 0 \,, \non  \\
R_{\pm}(\theta) &=& {1\over 2}(\cos {\xi\over 2} \pm i \sinh {\theta\over 2})
B(\theta - i \xi) B(\theta + i \xi) P(\theta) \,,
\label{boundS1}
\ee
where $P(\theta)$ is given by (\ref{Ptheta}). This set of amplitudes 
satisfies the supersymmetry constraints (\ref{SUSYconstraints1}),
with the parameter $\alpha$ which appears in the supersymmetry
charge (\ref{hatQ1}) given by
\be
\alpha= {i\over \cos {\xi\over 2}} \,.
\ee
We expect that the corresponding RG boundary flows should be given by 
Eq. (\ref{SUSYflows1}).

\section{Boundary TIM: R case}\label{sec:R}

We now wish to consider the perturbation of the tricritical Ising boundary CFT
with the CBC $(d)$. We have argued in Section \ref{sec:CBC} that the 
corresponding boundary state has superconformal symmetry. Hence, this
CBC is in fact superconformal,
\be
\left( T(y+ix) - \bar T(y-ix) \right) \Big\vert_{x=0} = 0 \,, \qquad 
\left( G(y+ix) + \bar G(y-ix) \right) \Big\vert_{x=0} = 0 \,.
\label{SCBC2}
\ee
Since the CBC $(d)$ is related to the CBC $(-0) \& (0+)$ by a duality
transformation (see Eq.  (\ref{duality})), we expect that the action
should be given by the image of the action (\ref{boundaryaction})
under duality. The bulk perturbing operator is invariant under duality.
(See Appendix \ref{sec:appendix}.) However, the individual conformal boundary
conditions $(-0)$ and $(0+)$ transform in a complicated manner under
duality, so that only their sum transforms simply.  Hence, the
boundary perturbing operator
${\cal D} \left( \phi_{({3\over 5})\,, (-0)}(y)
- \phi_{({3\over 5})\,, (0+)}(y) \right)$ 
is also complicated; in particular, it breaks spin-reversal symmetry,
and is different from the operator $\phi_{({3\over 5})\,, (d)}(y)$ considered 
in \cite{Ch}. The expected RG boundary flow is dual to (\ref{SUSYflows1}), 
namely,
\be
(d) & \rightarrow & (-) \& (+) 
\qquad \mbox{for} \qquad h<0 
\,, \non \\
(d) & \rightarrow & (0)
\qquad \qquad \mbox{for} \qquad h>0 \,.
\label{SUSYflows2}
\ee

By duality, we expect that the supersymmetry charge
\be
\hat Q = Q - \bar Q + {2 i \sqrt{m}\over \alpha} \Gamma
\label{hatQ2}
\ee
should be conserved.  This charge differs from the one considered
previously (\ref{hatQ1}) by the sign in front of $\bar Q$.
Requiring that this operator commute with the boundary $S$ matrix
(\ref{boundS}) yields the following constraints on the amplitudes:
\be
R_{+}(\theta) &=& R_{-}(\theta) \equiv R(\theta) \,, \non  \\
P_{+}(\theta) &=& P_{-}(\theta) \equiv P(\theta) \,, \non  \\
P(\theta) &=&  {i\over 2 \alpha \sinh {\theta\over 2}}
\left( V_{-}(\theta) - V_{+}(\theta) \right)\,.
\label{SUSYconstraints2}
\ee
These constraints are also a special case of those obtained in \cite{Ch} 
from the boundary Yang-Baxter equations.

We now proceed to determine the boundary $S$ matrix.  For $h=0$, all
three vacua are degenerate at the boundary, $e_{-1} = e_{0} = e_{1}$. 
We assume that for positive $h$, $e_{0} < e_{-1}=e_{+1}$; that is, the
state $0$ is the ground state of the
boundary, and the states $-1$ and $+1$ are degenerate excited states. 
We further assume that these excited states can be regarded as
boundary bound states associated with a pole of $R_{\pm}(\theta)$ 
at $\theta = i \xi$ which lies in the physical strip. 
The energies of the three states are therefore related by
\be
e_{\pm 1} = e_{0} + m \cos \xi \,,
\ee
implying that $h=0$ corresponds to $\xi = {\pi\over 2}$.
Moreover, for $\theta \sim i \xi$,
\be
R_{\pm}(\theta) \sim {g_{0 \pm} g_{\pm 0} \over \theta - i \xi} 
\label{polebehavior}
\ee
where $g_{\pm 0}$ and $g_{0 \pm}$ are particle-boundary coupling
constants \cite{GZ}.  From the constraint $R_{+}(\theta)=
R_{-}(\theta) \equiv R(\theta)$ in Eq.  (\ref{SUSYconstraints2}), it
follows that
\be
r \equiv {g_{0-}\over g_{0+}} =  {g_{+0}\over g_{-0}} \,.
\ee
The boundary bound state bootstrap equations imply that
\be
P(\theta) = \left[ A_{0}(\theta - i \xi) A_{0}(\theta + i \xi) +
A_{1}(\theta - i \xi) A_{1}(\theta + i \xi) \right] R(\theta) 
\ee
and
\be 
V_{+}(\theta) &=& r \left[ A_{0}(\theta - i \xi) A_{1}(\theta + i \xi) +
A_{1}(\theta - i \xi) A_{0}(\theta + i \xi) \right] R(\theta) \non \\
V_{-}(\theta) &=& {1\over r} \left[ A_{0}(\theta - i \xi) A_{1}(\theta + i \xi) +
A_{1}(\theta - i \xi) A_{0}(\theta + i \xi) \right] R(\theta)\,.
\ee
Apart from the factors of $r$ in $V_{\pm}(\theta)$, these results 
coincide with those obtained in \cite{Ch}. It follows that
\be
R(\theta) &=& e^{-2 \gamma \theta} P_{\xi}^{CDD}(\theta)\ P_{min}(\theta) 
\,, \non \\ 
P(\theta) &=& \cos {\xi\over 2} A(\theta - i \xi) A(\theta + i \xi) R(\theta)
\,, \non \\ 
V_{+}(\theta) &=& -i r \sinh {\theta\over 2} 
A(\theta - i \xi) A(\theta + i \xi) R(\theta) 
\,, \non \\ 
V_{-}(\theta) &=& {-i \over r } \sinh {\theta\over 2} 
A(\theta - i \xi) A(\theta + i \xi) R(\theta) 
\,.
\label{boundS2}
\ee
This set of amplitudes 
satisfies the supersymmetry constraints (\ref{SUSYconstraints2}),
with the parameter $\alpha$ given by
\be
\alpha= {1-r^{2}\over 2 r \cos {\xi\over 2}} \,.
\label{alpha2}
\ee

For the opposite sign of $h$, the situation is reversed: 
$e_{-1} = e_{+1} < e_{0}$, and so the ground
state of the boundary is two-fold degenerate, consisting of the vacua
$-1$ and $+1$, and the vacuum $0$ is an excited state. Indeed,
the above boundary $S$ matrix seems to be consistent with the
boundary flows proposed in Eq.  (\ref{SUSYflows2}).

We have seen that the parameter $r$ is not determined by the
constraints of $S$-matrix theory (unitarity, crossing, etc.),
integrability, or supersymmetry.  It is clear that for $h=0$, the
action has spin-reversal symmetry (i.e., it commutes with $\Gamma$),
which implies $V_{+} = V_{-}$.  That is, $r=1$ for $h=0$ (i.e., for
$\xi={\pi\over 2}$).  However, this constraint is rather mild, as it
can be satisfied in infinitely many ways, e.g. $r = \sin \xi$.  We
expect that through a more detailed analysis of the boundary
perturbation it should be possible to completely determine $r$, as
well as $\xi$, in terms of $h$.  However, we shall not pursue this
problem here.

Finally, it should be noted that the parameter $r$ can be set to unity
by an appropriate gauge transformation \cite{GZ} of the kink operators,
\be
K_{0 \,, 1}(\theta) \rightarrow e^{i \varphi} K_{0 \,, 1}(\theta) \,,
\qquad
K_{0 \,, -1}(\theta) \rightarrow e^{-i \varphi} K_{0 \,, -1}(\theta) \,,
\ee
which transforms the amplitudes $V_{\pm}(\theta)$ as
\be
V_{+}(\theta) \rightarrow e^{-2i \varphi} V_{+}(\theta) \,,
\qquad
V_{-}(\theta) \rightarrow e^{2i \varphi} V_{-}(\theta) \,,
\ee
and leaves the other amplitudes unchanged.  Ghoshal and Zamolodchikov
have argued that performing a gauge transformation corresponds to
adding a total derivative term to the boundary action density.  Thus,
setting $r \rightarrow 1$ corresponds to adding a total derivative
term to the boundary action which restores spin-reversal symmetry, in
which case the supersymmetry charge (\ref{hatQ2}) reduces to $\Gamma$,
since $\alpha \rightarrow 0$ (\ref{alpha2}).  It is this limiting case
which was considered in \cite{Ch}.

\section{Discussion}\label{sec:disc}

We have seen that it is possible to maintain both supersymmetry and
integrability in the boundary tricritical Ising model.  The NS case
(Section \ref{sec:NS}) corresponds to a ``direct sum'' of two
non-supersymmetric theories studied in \cite{Ch}.  The R case (Section
\ref{sec:R}) corresponds to a one-parameter ($r$) deformation of
another theory studied in \cite{Ch}.  For both cases, the conserved
supersymmetry charges (\ref{hatQ1}), (\ref{hatQ2}) are linear
combinations of $Q$, $\bar Q$ and the spin-reversal operator $\Gamma$.
For the other boundary supersymmetric integrable models which we have 
studied \cite{AN, Ne1}, the conserved supersymmetry charges have a 
similar structure. We expect that the phenomenon of forming 
superconformal boundary states from superpositions of pure Cardy 
states, which we have witnessed in the tricritical Ising model, may 
occur in other boundary supersymmetric models as well.

An important check on the picture presented here would be to verify
directly using the TBA that the proposed boundary $S$ matrices
(\ref{boundS1}), (\ref{boundS2}) describe the corresponding proposed
boundary flows (\ref{SUSYflows1}), (\ref{SUSYflows2}) .  This work is
now in progress \cite{AN2}.  It would also be interesting to clarify
the relation between the operator $\Theta$ in the perturbed CFT
expression (\ref{boundSUSY}) for the conserved supersymmetry charge,
and the operator $(-1)^{F}$, to which the spin-reversal operator
$\Gamma$ seems to correspond.

Our results should help to precisely determine the boundary $S$ matrix
for the solitons of the boundary $N=1$ supersymmetric sine-Gordon
model.  Indeed, this model is known to have supersymmetry, and the
breather boundary $S$ matrix has also been proposed \cite{Ne1}. 
However, the soliton boundary $S$ matrix \cite{AK}, which contains as
one of its factors the tricritical Ising model boundary $S$ matrix,
has not yet been completely determined.

We expect that ``higher'' perturbed boundary minimal models should
have generalizations of the boundary supersymmetry which we have
considered here.  For instance, the self-dual perturbation of the
tricritical 3-state Potts model (${\cal M}(6/7)$) has \cite{Za1} bulk
spin $1/3$ integrals of motion, certain combinations of which (plus
boundary terms) should be conserved in the corresponding integrable
boundary theory.  In general, perturbed minimal models have bulk
quantum group symmetries, which can survive in the presence of a
boundary.  It would be very interesting to work out the corresponding
conserved charges, boundary $S$ matrices and boundary flows.

\section*{Acknowledgments}

I am indebted to C. Ahn, J. Cardy, P. Dorey, K. Graham, R. Tateo and
A.B. Zamolodchikov for valuable discussions.  I am also grateful
for the hospitality extended to me at IPAM in UCLA and at the
University of Durham, where some of this work was done.  This work was
supported in part by the National Science Foundation under Grants
PHY-9870101 and PHY-0098088.

\appendix

\section{Duality}\label{sec:appendix}

The operator product algebra of the tricritical Ising CFT has a
discrete Kramers-Wannier-like symmetry known as duality \cite{Za1,
SCFT}.  We list here the duality transformation properties of some of
the primary fields.  The following primary fields are invariant under
duality:
\be
\Phi_{(0 \,, 0)} &\rightarrow &  \Phi_{(0 \,, 0)} \,, \qquad 
\Phi_{({3\over 5} \,, {3\over 5})} \rightarrow 
\Phi_{({3\over 5} \,, {3\over 5})} \,, \non \\
\Phi_{({1\over 10} \,, {3\over 5})} &\rightarrow & 
\Phi_{({1\over 10} \,, {3\over 5})} \,, \qquad 
\Phi_{({3\over 2} \,, 0)} \rightarrow  
\Phi_{({3\over 2} \,, 0)} \,.
\ee
The following primary fields pick up a minus sign under duality:
\be
\Phi_{({1\over 10} \,, {1\over 10})} &\rightarrow & 
-\Phi_{({1\over 10} \,, {1\over 10})} \,, \qquad 
\Phi_{({3\over 5} \,, {1\over 10})} \rightarrow  
-\Phi_{({3\over 5} \,, {1\over 10})} \,, \non \\
\Phi_{({3\over 2} \,, {3\over 2})} &\rightarrow & 
-\Phi_{({3\over 2} \,, {3\over 2})} \,, \qquad 
\Phi_{(0 \,, {3\over 2})} \rightarrow  
-\Phi_{(0 \,, {3\over 2})} \,.
\ee
The following primary fields pick up a factor of $G_{0}$ under 
duality:
\be
\Phi_{({3\over 80} \,, {3\over 80})}  \rightarrow   
G_{0} \Phi_{({3\over 80} \,, {3\over 80})} \,, \qquad 
\Phi_{({7\over 16} \,, {7\over 16})} \rightarrow  
G_{0} \Phi_{({7\over 16} \,, {7\over 16})} \,.
\ee 

\bigskip

\noindent
{\bf Note added:}

It was pointed out in \cite{KM} that the bulk $S$ matrix
(\ref{bulkamplitudes}) should be rescaled by a minus sign.  (See also
the correction of the scalar factor (\ref{bulkscalarfactor}).) 
Moreover, it was pointed out in \cite{MW} that the amplitude
$P(\theta)$ (\ref{Ptheta}) should be rescaled by the factor $i
\tanh({i \pi\over 4} - {\theta\over 2})$, in order that it have a
simple (rather than double) pole at $\theta= {i\pi\over 2}$ for $\xi =
{\pi\over 2}$.  Similarly, the amplitudes (\ref{boundS2}) should also
be rescaled by this factor.  I am grateful to L. Chim for bringing
\cite{MW} to my attention.

\vfill\eject 

\renewcommand{\theequation}{\arabic{equation}}
\setcounter{equation}{0}
\setcounter{footnote}{0}

\noindent    
{\large ERRATUM to ``Supersymmetry in the boundary tricritical Ising field theory''}

\bigskip

\noindent
{\bf Section 4: NS case}

\medskip 

We argued in \cite{ErNe} that the supersymmetric perturbation of the
``superposition'' of Cardy conformal boundary conditions $(-0) \&
(0+)$ is given by the boundary operator $\phi_{({3\over 5})\,,
(-0)}(y) -\phi_{({3\over 5})\,, (0+)}(y)$, i.e., with a relative minus
sign.  (See Eq.  (4.1).)  The argument is based on the results (see,
e.g., \cite{ErGRW}) for the ``pure'' boundary flows summarized in Eqs. 
(4.5), (4.6).  However, Graham has argued \cite{ErGr} that the signs
of the perturbing operators in (4.6) should be reversed; i.e., the
boundary flows are given by
\be
(-0) + \phi_{({3\over 5})} & \rightarrow &  (0) \,, \qquad 
(-0) - \phi_{({3\over 5})} \rightarrow (-) \,, \label{ErnonSUSYflows1} \\
(0+) - \phi_{({3\over 5})} & \rightarrow & (+) \,, \qquad 
(0+) + \phi_{({3\over 5})} \rightarrow (0) \,. \label{ErnonSUSYflows2}
\ee 
Therefore, in order to obtain the supersymmetric flows (4.7) of the
superposition of conformal boundary conditions $(-0) \& (0+)$, the
perturbing boundary operator should instead be
\be
\phi_{({3\over 5})\,, (-0)}(y) + \phi_{({3\over 5})\,, (0+)}(y) \,,
\label{Erplus}
\ee 
i.e., with a relative {\bf plus} sign. It follows that the boundary 
perturbation is {\bf even} under spin reversal, instead of odd. (The
arguments given in \cite{ErNe} that the perturbation is integrable and 
supersymmetric (4.3), (4.4) also hold if the relative sign is plus.)

We also argued in \cite{ErNe} that the boundary $S$ matrix for the NS
case, corresponding (e.g., for boundary parameter $h>0$) to the
boundary flow $(-0) \& (0+) \rightarrow (-) \& (+)$, should be given
by the ``direct sum'' of boundary $S$ matrices corresponding to the
two pure boundary flows $(-0) \rightarrow (-)$ and $(0+) \rightarrow
(+)$, which were found by Chim \cite{ErCh}.  However, the explicit
expression for the boundary $S$ matrix which we proposed there cannot
be strictly correct, since it is not consistent with the boundary
bound-state bootstrap equations \cite{ErGZ}.  \footnote{We were aware of
this difficulty, and we suggested in \cite{ErNe} that this constraint
should not be imposed.  The new boundary $S$ matrix which we propose
here does not involve such unphysical assumptions.} Moreover, it does
not have spin-reversal symmetry.

In order to formulate the ``direct sum'' of two boundary $S$ matrices, we
consider two ``copies'' of the boundary states.  Hence, instead of the
boundary operator $B_{a}$ which has a single index $a \in \{ -1\,, 0 \,,
1\}$ (corresponding to the three possible vacua or ``spins''), we now
have the boundary operator $B_{(a, \epsilon)}$ where $a
\in \{ -1\,, 0 \,, 1\}$ and $\epsilon \in \{ -1\,, 1\}$.
Boundary scattering is described by the boundary $S$ matrix $\R{(c,
\epsilon')}{a}{(b, \epsilon)}(\theta)$, which is defined by
\be
| K_{a,b}(\theta)\ B_{(b, \epsilon)} \rr_{in} =
\sum_{c, \epsilon'} \R{(c, \epsilon')}{a}{(b, \epsilon)}(\theta)\
| K_{a,c}(-\theta)\ B_{(c, \epsilon')} \rr_{out}
\,.
\ee
In particular, for the boundary flow $(-0) \& (0+) \rightarrow (-) \&
(+)$, we assume that the boundary can exist only in the following four
states, labeled by $(a, \epsilon)$: the two degenerate ground states
$(-1,-1)\,, (1,1)$ and the two degenerate excited states $(0,-1)\,,
(0,1)$.  Moreover, we propose that the boundary $S$ matrix has the
following nonvanishing amplitudes
\be
\R{(-1,-1)}{0}{(-1,-1)}(\theta) 
=\R{(1,1)}{0}{(1,1)}(\theta) &=& P(\theta) \,, \non \\
\R{(0,-1)}{1}{(0,-1)}(\theta) 
=\R{(0,1)}{-1}{(0,1)}(\theta) &=& R_{+}(\theta) \,, \non \\
\R{(0,-1)}{-1}{(0,-1)}(\theta) 
=\R{(0,1)}{1}{(0,1)}(\theta) &=& R_{-}(\theta) \,, 
\label{ErnewSmatrix}
\ee
where $P(\theta)$ is given by Eqs.  (4.16)-(4.19) in \cite{ErNe}; and
$R_{\pm}(\theta)$ are given by (4.21) when the boundary parameter
$\xi$ is in the range $0 < \xi < {\pi\over 2}$ (in which case the
simple pole of the amplitude $P(\theta)$ at $\theta = i\xi$
corresponds to a boundary bound state), and are otherwise zero.  The
amplitudes with $\epsilon=-1$ correspond to the boundary flow $(-0)
\rightarrow (-)$; while the amplitudes with $\epsilon=1$ correspond to
the boundary flow $(0+) \rightarrow (+)$.  One can verify that this
boundary $S$ matrix satisfies the boundary Yang-Baxter equation,
boundary unitarity, boundary cross-unitarity, as well as the boundary
bound-state bootstrap equations.

We now argue that this boundary $S$ matrix has both supersymmetry and
spin-reversal symmetry.  To this end, we assume that the supersymmetry
operators $Q$ and $\bar Q$ and the spin-reversal operator $\Gamma$
have the following action on the boundary operators for the ground
states:
\be
Q\ B_{(\pm 1, \pm 1)} = 0 = \bar Q\  B_{(\pm 1, \pm 1)} \,, \qquad
\Gamma\  B_{(\pm 1, \pm 1)} = B_{(\mp 1, \mp 1)} \,.
\ee 
Following Bajnok {\it et al.} \cite{ErBPT}, we determine the action of
these operators on the boundary operators for the excited states by
using a bootstrap construction for the latter, namely,
\be
B_{(0, \mp 1)} = {1\over g_{\mp 0}} K_{0, \mp 1}(i \xi) B_{(\mp 1,\mp 1)} \,,
\ee
where $g_{\mp 0}$ are particle-boundary coupling constants. Indeed,
it follows that 
\be
Q\ B_{(0, \mp 1)} = \pm \sqrt{m} e^{i \xi\over 2} B_{(0, \mp 1)} \,, 
\qquad
\bar Q\ B_{(0, \mp 1)} = \pm \sqrt{m} e^{-{i \xi\over 2}} B_{(0, \mp 1)} \,, 
\ee
and
\be 
\Gamma\ B_{(0, -1)} = r B_{(0, 1)}\,, \qquad 
\Gamma\ B_{(0, 1)} = {1\over r} B_{(0, -1)}\,, 
\ee
where $r=g_{+0}/g_{-0}$.  One can now verify that both $Q + \bar Q$
and $\Gamma$ commute with the boundary $S$ matrix (\ref{ErnewSmatrix}). 
These properties are consistent with those of the boundary
perturbation (\ref{Erplus}).  In particular, the operator
\be
\hat Q = Q + \bar Q + {2 i \sqrt{m}\over \alpha} \Gamma \,,
\ee 
(see (4.14) in \cite{ErNe}) commutes with the boundary $S$ matrix
for any value of $\alpha$.  (Hence, the restriction
(4.23) is not necessary.)  Following the arguments of \cite{ErBPT}, the
value of $\alpha$ is related to the energy of the ground state,
$E_{0}=-2m/\alpha^{2}$.

\bigskip

\noindent
{\bf Section 5: R case}

\medskip 

For the R case, the boundary perturbing operator is the dual of the
operator (\ref{Erplus}), which is given by $\phi_{({3\over 5})\,,
(d)}(y)$.  This operator is even under spin-reversal.  Precisely this
case was considered by Chim \cite{ErCh}.  The corresponding boundary
flows are given by (5.2).

For the boundary flow $(d) \rightarrow (0)$, the boundary has ground
state $0$, and degenerate excited states $\pm 1$.  As in \cite{ErNe}, we
assume that the supersymmetry operators $Q$ and $\bar Q$ and the
spin-reversal operator $\Gamma$ act on the boundary operator for the
ground state as follows,
\be
Q\ B_{0} = 0 = \bar Q\  B_{0} \,, \qquad \Gamma\  B_{0} =  B_{0} \,.
\ee
However, following Bajnok {\it et al.} \cite{ErBPT}, let us now
use a bootstrap construction for the boundary operators for the
excited states
\be
B_{\mp 1} = {1\over g_{0 \mp}} K_{\mp 1, 0}(i \xi) B_{0} 
\ee
to determine how the symmetry operators act on them:
\be
Q\ B_{1} &=&  i r\sqrt{m} e^{i \xi\over 2} B_{-1} \,, \qquad
Q\ B_{-1} = - {i\over r}\sqrt{m}   e^{i \xi\over 2} B_{1} \,,\non \\
\bar Q\ B_{1} &=& - i r\sqrt{m} e^{-{i \xi\over 2}} B_{-1} \,, \qquad
\bar Q\ B_{-1} =  {i\over r}\sqrt{m} e^{-{i \xi\over 2}} B_{1}\,, \non \\
\Gamma\ B_{1} &=& r B_{-1} \,, \qquad \Gamma\ B_{-1} = {1\over r} 
B_{1} \,,
\ee
where $r=g_{0-}/g_{0+}$. (In contrast, we assumed in \cite{ErNe} that
$Q$ and $\bar Q$ annihilate all the boundary operators $B_{a}$.)
Demanding that both $Q - \bar Q$ and $\Gamma$ commute with the
boundary $S$ matrix now leads to the following 
constraints on the amplitudes:
\be
R_{+}(\theta) &=& R_{-}(\theta) \equiv R(\theta) \,, \non  \\
P_{+}(\theta) &=& P_{-}(\theta) \equiv P(\theta) \,, \non  \\
r V_{-}(\theta) &=& {1\over r} V_{+}(\theta) \,, \qquad
P(\theta) = {i  r\cos {\xi\over 2} \over \sinh {\theta\over 2}} V_{-}(\theta)
\,.
\ee
Comparing with Eq.  (5.4) in \cite{ErNe}, we see that the first two
lines are the same, but the third line is different.  The boundary $S$
matrix (5.10) satisfies all of these constraints.  In particular, the
operator (5.3)
\be
\hat Q = Q - \bar Q + {2 i \sqrt{m}\over \alpha} \Gamma 
\ee 
commutes with the boundary $S$ matrix for any value of $\alpha$. 
(Hence, the restriction (5.11) is not necessary.)  Following the
arguments of \cite{ErBPT}, the value of $\alpha$ is related to the
energy of the ground state, $E_{0}=-2m/\alpha^{2}$.

I am grateful to C. Ahn, L. Chim and K. Graham for their valuable 
comments. This work was supported in part by the National Science
Foundation under Grant PHY-0098088.

\end{document}